\documentclass[twocolumn,floatfix]{aastex62}

\usepackage{amsmath}
\usepackage{epstopdf}
\usepackage{hyperref} 
\hypersetup{colorlinks,citecolor=blue,linkcolor=blue,urlcolor=blue}
\usepackage{url}
\graphicspath{{./figures/}}
\usepackage{lineno}
\usepackage{savesym}
\savesymbol{tablenum}
\usepackage{siunitx}
\restoresymbol{SIX}{tablenum}
\begin{document}

\title{Automated transient detection with shapelet analysis in image-subtracted data}

\author{Kendall Ackley}
\affiliation{Monash Centre for Astrophysics, School of Physics and Astronomy, Monash University, VIC 3800, Australia}
\affiliation{OzGrav: The ARC Centre of Excellence for Gravitational Wave Discovery, Clayton VIC 3800, Australia}
\email{kendall.ackley@monash.edu}
\author{Stephen S. Eikenberry}
\affiliation{Department of Physics, University of Florida, Gainesville, FL 32611, USA}
\affiliation{Department of Astronomy, University of Florida, Gainesville, FL 32611, USA}
\author{Ceren Yildirim}
\affiliation{Department of Astronomy, University of Florida, Gainesville, FL 32611, USA}
\author{Sergei Klimenko}
\affiliation{Department of Physics, University of Florida, Gainesville, FL 32611, USA}
\author{Alan Garner}
\affiliation{Department of Astronomy, University of Florida, Gainesville, FL 32611, USA}

\published{10 October 2019}
\submitjournal{AJ}
%\date{\today}

\begin{abstract}

We present a method for characterizing image-subtracted objects based on shapelet analysis to identify transient events in ground-based time-domain surveys. We decompose the image-subtracted objects onto a set of discrete Zernike polynomials and use their resulting coefficients to compare them to other point-like objects. We derive a norm in this Zernike space that we use to score transients for their point-like nature and show that it is a powerful comparator for distinguishing image artifacts, or residuals, from true astrophysical transients. Our method allows for a fast and automated way of scanning overcrowded, wide-field telescope images with minimal human interaction and we reduce the large set of unresolved artifacts left unidentified in subtracted observational images. We evaluate the performance of our method using archival intermediate Palomar Transient Factory and Dark Energy Camera survey images. However, our technique allows flexible implementation for a variety of different instruments and data sets. This technique shows a reduction in image subtraction artifacts by 99.95\,\% for surveys extending up to hundreds of square degrees and has strong potential for automated transient identification in electromagnetic follow-up programs triggered by the Laser Interferometer Gravitational Wave Observatory-Virgo Scientific Collaboration.

\end{abstract}


\section{Introduction}

A new generation of ground-based optical instruments specializing in
wide-field, high-cadence, deep imaging transient surveys such as Palomar Transient Factory \citep[PTF;][]{Law2009,Rau2009}, Dark Energy Camera \citep[DECam;][]{Flaugher2015}, SkyMap\-per \citep{Keller2007}, Catalina Real-Time Transient Survey \citep[][]{Drake2009}, LaSilla Quest \citep[][]{Rabinowitz2012}, Hyper SuprimeCam \citep[][]{Miyazaki2012}, MegaCam \citep{Boulade2003},  Asteroid
Terrestrial-impact Last Alert System \citep[ATLAS;][]{Tonry2011}, Panoramic Survey Telescope and Rapid Response System \citep[Pan-STARRS;][]{Kaiser2010}, All-Sky Automated Survey for Supernovae  \citep[ASAS-SN;][]{Shappee2014}, the Gravitational-Wave Optical Transient Observer (GOTO \footnote{\url{http://goto-observatory.org}}), BlackGEM \citep{Bloemen2015}, Evryscope \citep[][]{Law2014} and Zwicky Transient Facility \citep{Bellm2019}, or the upcoming Large Synoptic Survey Telescope \citep[LSST,][]{Ivezic2019}, are opening a new window for transient astronomy. The \`{e}ntendue of these surveys are expected to generate an exponential increase in data collection from what is currently generated, necessitating fully automated processing and vetting of transient candidates.

We focus on developing a technique that retains real astrophysical transients such as supernovae (SNe), gamma-ray burst (GRB) orphan afterglows, active galactic nuclei (AGN), blazars, dwarf novae, and other potential transient events such as tidal disruptions\,\citep{Strubbe2009} or shock breakouts\,\citep{Garnavich2016} while reducing or eliminating the number of image subtraction residuals (i.e. non-astrophysical artifacts) which would otherwise litter the image in vast quantities over any given observing time frame. One compe-ling motivation for the development of this technique is the non-trivial task of associating an electromagnetic (EM) transient with a gravitational-wave (GW) candidate with both angular and time coincidence. True GW-associated transients will be rare in our imaging field so the challenge at hand is to efficiently remove unassociated steady-state image subtraction residuals and further reduce the number of events from image subtracted data using our method.

A majority of the steady-brightness objects that are retained between exposures can be removed by image differencing techniques. However, image subtraction artifacts are expected to comprise a large fraction of the possible transient candidates to be vetted and can be as high as hundreds to thousands per image. Thus, the sheer number of detected objects makes manual processing impractical, and automatic pre-filtering of candidate transient objects becomes necessary. Besides a look-up for known variable sources that may be in the telescope's field of view, any automated selection mechanism should include a test for the point-like nature of an object.

There are several methods which address the issue of artifact reduction and transient classification in an automated, machine-learned environment for many optical transient surveys, nearly all of which rely on the use of a human-classified training set of data. \cite{Bloom2012} and \cite{Brink2013} introduce a supervised machine-learning real--bogus algorithm using a random forest (RF) classifier to rank cataloged transients in PTF data, while\,\cite{Wright2015} expand on the real--bogus algorithm for Pan-STARRS1 confirming the successful performance of RF over the artificial neural network (ANN) and support vector machine (SVM) algorithms. \cite{Bailey2007} explore SVM, RF, and boosted decision trees (BDT) for the supernova classification pipeline Nearby Supernova Factory\,\citep{Aldering2002}, showing that BDT provides the best overall performance. \cite{Goldstein2015} for the Dark Energy Survey Supernova Program \citep[DES-SN;][]{Bernstein2012} apply RF classification to their data set, while \cite{Morii2016} compare RF, the deep neural network, and boosting based on maximizing the area under the receiver-operating characteristic (ROC) curve (AUC) for HSC data and show the increased utility in combining all three. \citet{Klencki2016} implement a neural network unsupervised algorithm self-organizing map for OGLE-IV\,\citep{Wyrzykowski2014}. A large comparison of machine-learning techniques \,\citep[see][]{duBuisson2015} shows that RF outperforms all other supervised methods. While many of these techniques may employ similar methods, such as generalized feature engineering and the choice of classification algorithms, there is no set benchmark data in order to directly compare any of these methods to one another.

Our method makes use of singular value decomposition (SVD) for learning latent variables. We provide an initial training set of unsubtracted images with labels given by the \texttt{CLASS\_STAR} parameter from \texttt{SExtractor}\,\citep{BertinArnouts1996}. However, we do not human-classify any source in the images. Instead, we have code automatically define the shape of a point source from individual images and score any point-like objects that appear in the final difference image using a metric in a Zernike space. This idea is applicable to many individual telescopes, without the need for labor-intensive manual training of any algorithms and is distinct from traditional machine-learning methods. This can be particularly useful at the onset when new instruments have not collected enough training data to enable finding such transients during their observations, or for facilities which do not have the resources to manually train their algorithms.

We discuss the general method of image subtraction in \S\ref{imagesubtraction}. We discuss our Zernike method in \S\ref{zernike} including a general description of point-spread function (PSF) modeling (\S\ref{psf}), the set of Zernike polynomials (ZPs; \S\ref{zps}), our decomposition method (\S\ref{decomp}) and parameters (\S\ref{params}), the importance of subpixel shifting (\S\ref{shift}), and the metric we use for our analysis (\S\ref{norm}). In \S\ref{results} we detail our method of injecting artificial transients and show our results in terms of efficiency and false alarm rate (FAR). We provide discussion and compute local system run times in \S\ref{conclusions} and \S\ref{runtime}. We conclude in \S\ref{conclusions}.

\section{Image Subtraction}
\label{imagesubtraction}

Difference imaging is a highly effective method for detecting and obtaining time-series photometry of transient events in crowded regimes. It is more efficient than PSF aperture photometry at extracting information of a transient or variable star's light curve over time. 

The basic method of difference imaging subtracts two exposures from the same field to suppress steady-state brightness objects. \citet[][hereafter AL98]{AlardLupton1998} developed a technique for optimal image subtraction released as part of their \texttt{ISIS} package \citep{AlardLupton1998,Alard2000}. The AL98 method subtracts the two images by finding a convolution kernel that remaps the to-be-subtracted image into the coordinate frame of the primary frame and matches the seeing. The kernel $K$ for subtracting the reference image $R$ from the observational image $I$ is obtained by minimizing the quadratic error in the expression of the function
\begin{equation}
R(x,y) \ast K(x,y) - I(x,y) - B(x,y),
\end{equation} 
where the slowly varying $B$ accounts for any background variations present between the two images. 

In the context of wide-field images, image subtraction is the preferred method for identifying residual energy between two exposures and will ideally remove all steady-state sources. However, in reality subtraction artifacts are introduced due to saturation effects or from variable PSFs, with the latter being particularly important in images with coverage on the order of square degrees. Image artifacts manifest due to the inherent challenge of finding an optimal kernel solution especially when significant seeing variations are present. A global optimal kernel solution for the image may not fully account for the deviations of the PSF around extended (galaxy) sources and convolution will lead to over-subtraction especially if the observation image is found to have better seeing than the reference. To mitigate the effects resulting from seeing differences, we can perform image subtraction in both directions.

There are several subtraction algorithms available that will identify a cross-convolution kernel and transform two images into a common frame, where objects that appear in both with the same intensity will cancel out upon differencing\,\citep{Bramich2008, Kessler2015, Zackay2016}. The algorithms each have their own strengths and shortcomings, however, for this analysis we use the subtraction tools included in the \texttt{HOTPANTS} \citep{Becker2015} program, which is based on the AL98 algorithm but offers more flexibility in the subtraction process, particularly in its ability to choose the direction of subtraction.

As the subtracted image is in the same frame as the observational image $I$, the PSF will be identical where the presence of background fluctuations between the two images is removed during image subtraction. For this reason we build a PSF library based on sources within $I$ and require that the reference image $R$ is used solely for subtraction. 

\section{Zernike Decomposition}
\label{zernike}
Analytically approximating the wavefronts in the aperture frame using expansions of ZPs is a popular method for optics analysis via monitoring incoming wavefronts\,\citep{Lofdahl1994} and in the field of adaptive optics\,\citep{Rigaut1991} where dynamically adjusting mirrors automatically corrects for incoming aberrated wavefronts. They are also incorporated outside of the field of astronomy, specifically in image pattern recognition\,\citep{Khotanzad1990}. We make use of ZPs as our basis set to decompose the image objects. 

\subsection{PSF Modeling and Shapelet Analysis}
\label{psf}

The PSF of an astronomical image is the intensity distribution produced by a true point source at the sensor and is affected by static optical aberrations of the telescope/instrument as well as time-varying atmospheric effects, e.g. seeing. Near the center of the optical axis, a simple 2D Gaussian may be sufficient for fitting in some case. However, as objects deviate from the center, or more complex turbulence profiles occur, models with higher-dimensional solutions must typically be used for fitting. One way to address the abberated PSFs is to use an interpolated PSF from actual data. 
A fit to the PSF with amplitude and position as fitting parameters resembles the ultimate test of the point-like characteristic of an object found in the image. However, the PSF is not known \emph{a priori} and must be empirically determined.

There are many base sets which may be appropriate for accurately deconvolving and fitting a deviated wavefront in the aperture frame, such as shapelets: Zernike \citep{Conforti1983}, Hermite--Gauss \citep{Refregier2003}, and Gauss--Laguerre polynomials \citep{MasseyRefregier2005}; cheblets \citep{JimenezTeja2012}; or wavelets \citep{Starck2003}. \cite{Piotrowski2013} consider a modified shapelet basis using a set of ZPs for modeling wide-field PSFs as a function of distance from the aperture. We revise traditional techniques of PSF decomposition and modeling by looking instead at individual objects in the image plane, based on the premise that point-like sources will retain the characteristic properties of the global image PSF. While we make use of ZPs for our particular study, the current structure of our pipeline allows for interchangeable basis functions to fit individual needs or preferences.

The shapelet analysis is largely model-independent in that it is less prone to errors than for a single distribution, i.e. a Gaussian. For example, in choosing a Gaussian model to fit the PSF there may be significant errors for higher-order deviations. If instead an entire basis set is chosen, e.g. Zernike or Hermite--Gauss polynomials, such deviations can be effectively handled by using the higher-order moments.

\subsection{Zernike Polynomials}
\label{zps}
The ZPs in their analytical form are a complete orthonormal basis for real-valued, smooth functions on the unit disk $(0\leq\rho\leq1,$ $0\leq\phi<2\pi)$. They are grouped into even and odd polynomials, depending on their symmetry in $\phi$, and are defined as
\begin{eqnarray}
Z^{m}_{n}\left(\rho,\phi\right)&=&R^{m}_{n}\left(\rho\right)\cos\left(m\phi\right),\hspace{.2in}\mathrm{and}\\
Z^{-m}_{n}\left(\rho,\phi\right)&=&R^{m}_{n}\left(\rho\right)\sin\left(m\phi\right),
\end{eqnarray}
respectively, where
\begin{equation}
R^{m}_{n}\left(\rho\right) = \sum^{\left(n-m\right)/2}_{k=0}\frac{\left(-1\right)^k\left(n-k\right)! \rho^{n-2k}}{k!\left(\frac{n+m}{2} -k \right)!\left(\frac{n-m}{2} -k \right)!}.
\end{equation}
The primary index $n\in\{0,1,2,\ldots\}$ determines the order of the polynomial in $\rho$, and $m\in\{n,n-2,n-4,\ldots\}$ is the axial index.
\begin{figure}[t]
\includegraphics[width=\columnwidth]{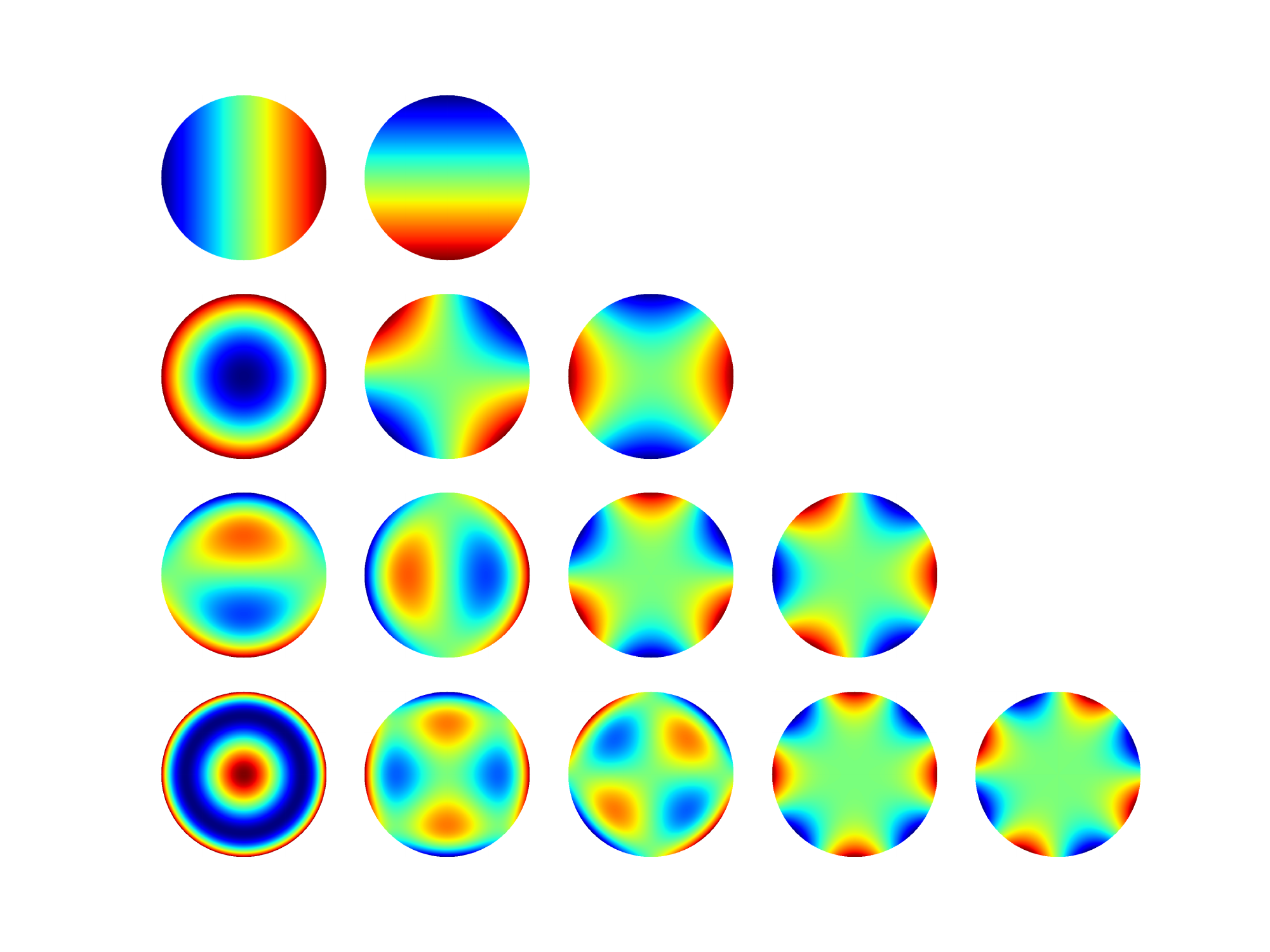}
\caption{Example of the first 14 Zernike Orders.}
\label{fig:zern_pol_ex}
\end{figure}
Fig.\,\ref{fig:zern_pol_ex} shows a few exemplary low-order ZPs. A smooth function $f(\rho,\phi)$ can be written as the infinite series
\begin{equation}
f(\rho,\phi)=\sum_{m,n}\left[a_{m,n}Z^m_n(\rho,\phi)+b_{m,n}Z^{-m}_n(\rho,\phi)\right],
\label{eq:f_coeffs}
\end{equation}
where $a_{m,n}$ and $b_{m,n}$, are the respective symmetric and antisymmetric coefficients of the Zernike expansion. Due to the ZP's orthonormality, they can be calculated from
\begin{equation}
\label{eq:Z_coeff_calc}
	\begin{aligned}
a_{m,n}&=\int_{0}^{1}\limits\int_{0}^{2\pi}\limits f(\rho,\phi) Z_n^{m}(\rho,\phi)\,d\phi\,d\rho \\
b_{m,n}&=\int_{0}^{1}\limits\int_{0}^{2\pi}\limits f(\rho,\phi) Z_n^{-m}(\rho,\phi)\,d\phi\,d\rho .
	\end{aligned}
\end{equation}
Because of their importance in describing wavefront aberrations, some particular index pairs have been given names, i.e. astigmatism ($a_{2,2}$ and $b_{2,2}$) or coma ($a_{1,3}$ and $b_{1,3}$). In this paper we use the Noll-ordering numbering scheme\,\citep{Noll} for ZPs that only uses a single index ($Z_n^m \rightarrow Z_j$).

Choosing ZPs as a basis for the residual representation is met with several complications. Firstly, there is the discretization error that inevitably results from the numerical evaluation of the ZPs on a pixel grid of finite resolution. Secondly, the PSF is theoretically infinite in expanse in the image plane, while the ZPs are defined only within an aperture. An appropriate unit disk size on the pixel grid must therefore be chosen such that only a trace amount of PSF power remains outside the disk, while still being computationally feasible. Lastly, and most importantly, for the exact representation of an arbitrary function, even in the analytical case, infinitely many Zernike coefficients are required. Computationally this is impossible and a cut-off criterion for the ZP order must be developed. Both aperture restriction and order limitation need to be weighed against the numerical error that they introduce. It should be noted that the larger the disk, the higher the Zernike orders must be in order to spatially resolve PSF features to the same accuracy.

\begin{center}
\begin{figure*}[t]
\includegraphics[width=\textwidth]{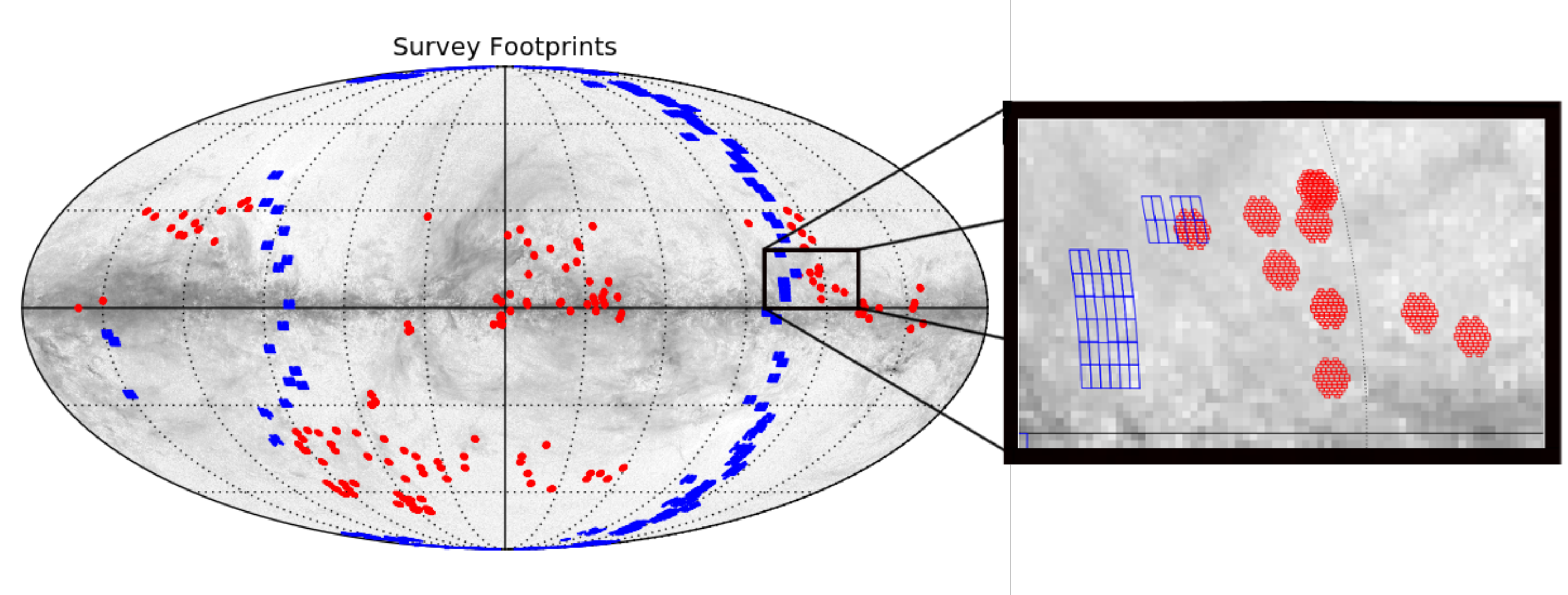}%
\caption{Collection of iPTF $R$-band DR1 and DR2 (blue) and DECam-DESY1 (red) survey images projected onto a Mollweide axis in ecliptic coordinates. The inset image shows detailed coverage and decl. overlap.}%
\label{fig:ptf_survey}%
\end{figure*}
\end{center}

\subsection{Decomposition Method}
\label{decomp}
The purpose of analytical Zernike decomposition is to find coefficients $c_j$ corresponding to $a_{m,n}$ and $b_{m,n}$ in Eq.\,\eqref{eq:f_coeffs} and reconstruct the object $f_{\mathrm{obj}}[k,l]$ in the basis of the ZPs as
\begin{equation}
\hat{f}_{\mathrm{obj}}[k,l]=\sum_jc_jZ_j[k,l],
\label{eq:rec_wf}
\end{equation}
where $k$ and $l$ are the row and column indices of the pixel grid, respectively. For this an aperture is placed around the object to be decomposed. To accurately capture the contribution of edge pixels, which may only partially overlap with the unit disk, the aperture condition $\rho\leq1$ is evaluated on a much finer grid than that of the image, i.e. the grid is up-sampled by a factor of 10. The edge pixels are then averaged and down-sampled to match the original image pixel grid, yielding a mask $p[k,l]$ whose values give the fractional coverage of pixel $[k,l]$ by the up-sampled unit disk. Normalization of the apertured object yields the distribution function $f_{\mathrm{obj}}[k,l]$. Similarly, the ZPs are first calculated on the higher precision grid, and then averaged and down-sampled to yield the basis $Z_j[k,l]$ for the source decomposition.

Due to the finite number of ZPs used in the decomposition process, an exact reconstruction of $f_{\mathrm{obj}}[k,l]$ is typically not possible. We therefore introduce the requirement that the Zernike representation of the reconstructed object must not exceed a threshold value of the merit function
\begin{equation}
\sigma_{f} = \sqrt{\frac{\sum_{k,l}\limits\left[p[k,l]\left(f_{\mathrm{obj}}[k,l] -\hat{f}_{\mathrm{obj}}[k,l]\right)^2\right]}{\sum_{k,l}\limits p[k,l]}}
\label{eq:rms_err}
\end{equation}
which is the rms deviation between the reconstructed object $\hat{f}_{\mathrm{obj}}[k,l]$ and the original $f_{\mathrm{obj}}[k,l]$ on the mask $p[k,l]$. To obtain the particular set of $c_j$ that meets this requirement we must first calculate the Zernike moments
\begin{equation}
M_j=\sum_{k,l}\Big(f_{\mathrm{obj}}[k,l]\times Z_j[k,l]\Big)
\label{eq:Z_mom_calc}
\end{equation}
analogously to \eqref{eq:Z_coeff_calc}. With the inverse $\mathcal{C}_{ij}^{-1}$ of the covariance matrix
\begin{equation}
\mathcal{C}_{ij}=\sum_{k,l}\Big(Z_i[k,l]Z_j[k,l]\Big),
\label{eq:covariance_matrix}
\end{equation}
which is calculated using SVD and Gram--Schmidt orthogonalization, we obtain the $c_j$ as the inner product of $M_j$ with $\mathcal{C}_{ij}^{-1}$ from
\begin{equation}
c_j=\sum_iM_i\mathcal{C}_{ij}^{-1}.
\label{eq:red_Z_mom}
\end{equation}
In this discussion we use the discrete Zernike coefficients $c_j$ to characterize point sources in Zernike space.

\subsection{Decomposition Parameters}
\label{params}
Before a sophisticated efficiency study can be launched, some fundamental parameters for the decomposition process need to be chosen. As mentioned previously, there is a trade-off between the physical aperture radius $\rho_{\mathrm{pix}}$ and the Zernike cutoff order $j_{\mathrm{max}}$ for the ability to resolve wavefront features. Since Eq.\,\eqref{eq:rms_err} provides a measure for the quality of the Zernike fit, it served as our main criterion to determine pairs of $(\rho_{\mathrm{pix}},j_{\mathrm{max}})$ that lead to computationally feasible, yet accurate representations.

We vary $\rho_{\mathrm{pix}}$ from 3.0 to 7.0\,pixels. Motivated by the structure of the Noll ordering, we chose a set of $j_{\mathrm{max}}$, in which we parameterize our analysis. We note that $j_{\mathrm{max}}$, in general, should not exceed the number of valid pixels in $p[k,l]$.

We use a selection of archival intermediate Palomar Transient Factory \citep[iPTF;][]{Law2009,Rau2009} Data Release 1,2, and 3 (DR1, DR2, DR3) and Dark Energy Camera \citep[DECam][]{Flaugher2015} Dark Energy Survey Year 1 (DESY1) images of varying quality in terms of background and seeing. These images are publicly available and were imaged between 2009 September--2015 January and 2013 February--2014 December, respectively. The collection of images for iPTF and DECam is shown in Fig.\,\ref{fig:ptf_survey}. 
Auxiliary catalog files, which are generated by \texttt{SExtractor}, contain basic astrometric and photometric estimates, and supply locations and aperture magnitudes, as well as data quality flags for objects in the image.
We restrict our sources for the PSF model to isolated, unsaturated objects with minimum flux values $>50\%$ above the sky background.

\begin{figure}[t]%
	\includegraphics[width=\columnwidth]{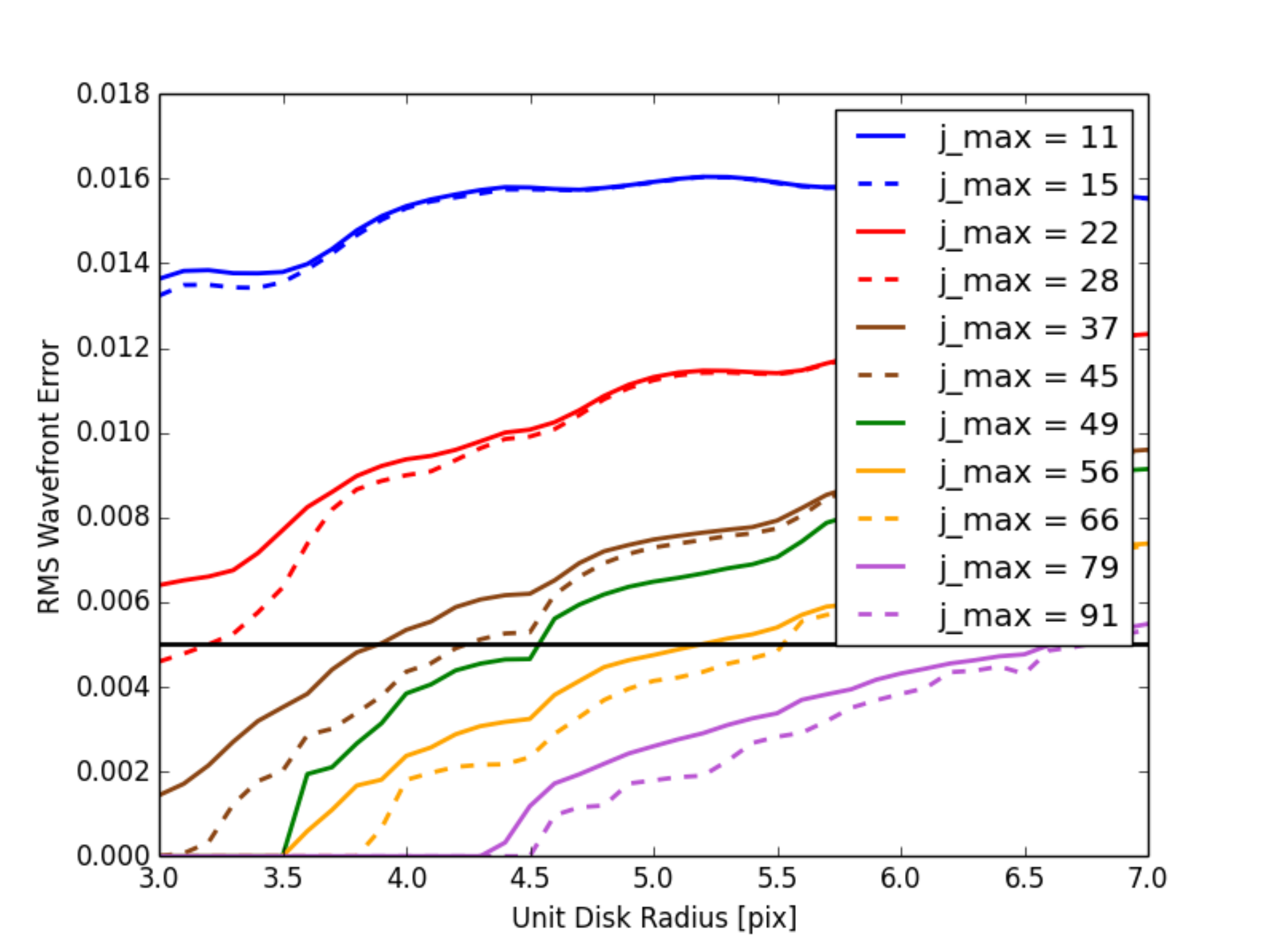}%
	\caption{Median RMS error for Zernike reconstruction of an ensemble of reference stars. We varied the unit disk radius $\rho$ from 3.0 to 7.0 pixels for different $j_{\mathrm{max}}$ and calculated the median residual error between the original and the Zernike-reconstructed intensity distributions over an ensemble of roughly 4000 objects. For an expected error of 0.5\% we decided to use $\rho_{\mathrm{disk}}=4.5$ for iPTF ($\rho_{\mathrm{disk}}=6.5$ for DECam) and $j_{\mathrm{max}}=49$ for our simulations.}
	\label{fig:rms_err}
\end{figure}

Using the source's coordinates we define a region with dimensions three times the corresponding unit disk centered on the object and calculate its median pixel value, which serves as an estimate for the nominal local background. We then apply the aperture $p[k,l]$ to the region and normalize the obtained intensity distribution, followed by the previously described Zernike decomposition. We calculate the residual error of the reconstructed sources following Eq.\,\eqref{eq:rms_err} for all objects that have passed a preselection filter. As a figure of merit for a pair $(\rho_{\mathrm{pix}},j_{\mathrm{max}})$ we use the median error across the ensemble of decomposed objects. Fig.\,\ref{fig:rms_err} shows the result of our analysis. We find that to keep the reconstructed error below 0.5\% with a moderately high $j_{\mathrm{max}}$ of 49, we set $\rho_{\mathrm{pix}}$ to 4.5 for iPTF and $\rho_{\mathrm{pix}}$ to 6.5 for DECam.

\subsection{Subpixel Shifting}
\label{shift}
\texttt{SExtractor} estimates the coordinates of the object with fractional pixel accuracy while the PSF is reconstructed with integer pixel accuracy, leading to off-axis PSF representations and a large error. One possible way to address this is to locally resample the telescope image before the Zernike decomposition and shift it by the fractional pixel value. However this modifies the observational data. Another option, which is the method we use, is to evaluate the ZPs with respect to an offset origin given by the object's coordinates estimate. The latter is more precise, but at the same time more computationally demanding, as it requires the recalculation and inversion of the covariance matrix for each shift.

\subsection{Zernike Statistics}
\label{norm}
Our declared goal is to develop a criterion for the point-like nature of an object using the Zernike approach. However we do not know the expected values of the Zernike coefficients of a point source initially, but must derive them using ensemble statistics. Our strategy is to compile a selection of reference stars from the \texttt{SExtractor} catalog and decompose each one individually. For each Zernike index $j$ we can determine a statistical mean $\bar{c}_j$ and a characteristic spread $\sigma_j$, which we use to define the distance from the ideal point source in Zernike space
\begin{equation}
D_{Z} = \sum_{j=1}^{j_{\mathrm{max}}}\frac{(c_j-\bar{c}_j)^2}{\sigma_j^2}
\label{eq:z_dist_def}
\end{equation} 
for an object with Zernike coefficients $c_j$. We use $D_Z$ as the main discriminator for the point-like shape of decomposed objects. This is also defined to be the main feature we use to distinguish between real and bogus sources in the subtracted image. We find that the discriminative behavior of $D_Z$ is relatively stable (weighing the appropriate observing conditions) and is tunable for each instrument. To decide which items from the catalog to use for obtaining $\bar{c}_j$ and $\sigma_j$ we use the \texttt{CLASS\_STAR} field which is a \texttt{SExtractor} estimate for the point-likeness of an object ranging from 0 (extended source) to 1 (point-like). We set 0.90 as a lower bound for \texttt{CLASS\_STAR}. Under the premise that \texttt{SExtractor} is correct in its guess for the majority of cases, we obtain realistic $\bar{c}_j$ and $\sigma_j$ through the averaging process.

\begin{figure}[t]
\includegraphics[width=\columnwidth]{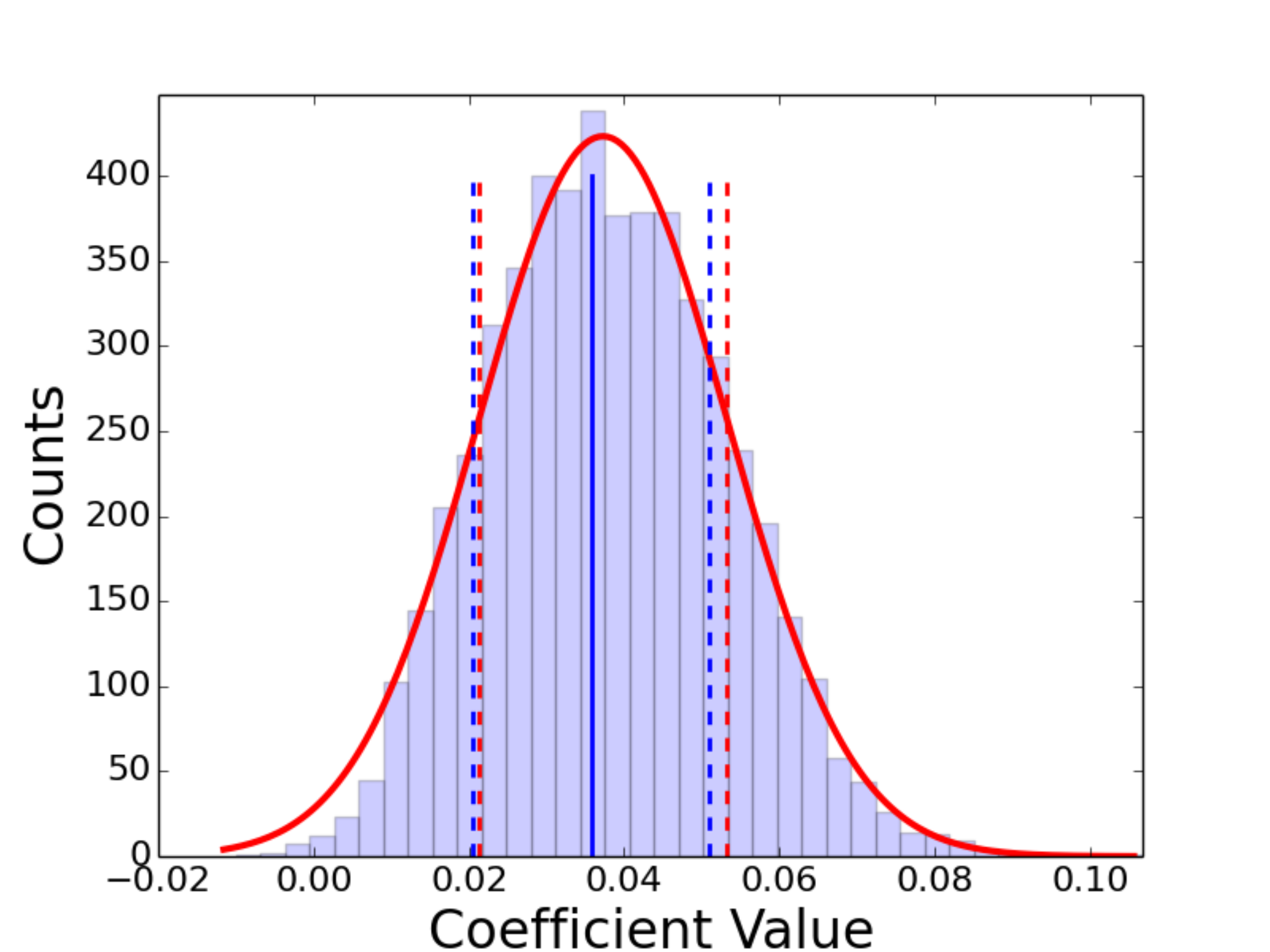}%
\caption{Example Gaussian distribution for a set of coefficients $c_j$ for a single Zernike order. The solid red line is a Gaussian fit to the data and the solid blue line is the ensemble median. The red and blue dashed lines show the median $\sigma_j$ and the FWHM of the Gaussian fit, respectively. Each Zernike order can be represented in this manner and the median coefficient for each order is taken from the median of the distribution.}
\label{fig:coeff_gaussian}
\end{figure}

We further apply filtering parameters and decompose the resulting selection of objects. Fig.\,\ref{fig:coeff_gaussian} illustrates a histogram of a $c_j$ distribution that is obtained as explained. The ensemble median and standard deviation, as well as a Gaussian fit to the presented data, have been included in the figure. The median and standard deviation are empirical values for $\bar{c}_j$ and $\sigma_j$. While the coefficient distribution are approximately Gaussian, using the median and standard deviation has proven to be more robust. As the seeing conditions vary between exposures these quantities are bound to change as well and a recalculation of $\bar{c}_j$ and $\sigma_j$ is required for each image.

The further the coefficient values are from their statistical mean, the less likely it is that a decomposed object is a point source. Collapsing all Zernike orders into the single metric from Eq.\,\eqref{eq:z_dist_def} we develop a threshold value for $D_Z$ as the main criterion of our pipeline. The exact value will need to be chosen weighting the pipeline's false positive rate (FPR) against its detection efficiency and is a tunable parameter for each telescope.

\section{Method Testing and Results}
\label{results}

\begin{figure*}[t]
\centering
\includegraphics[width=0.9\linewidth]{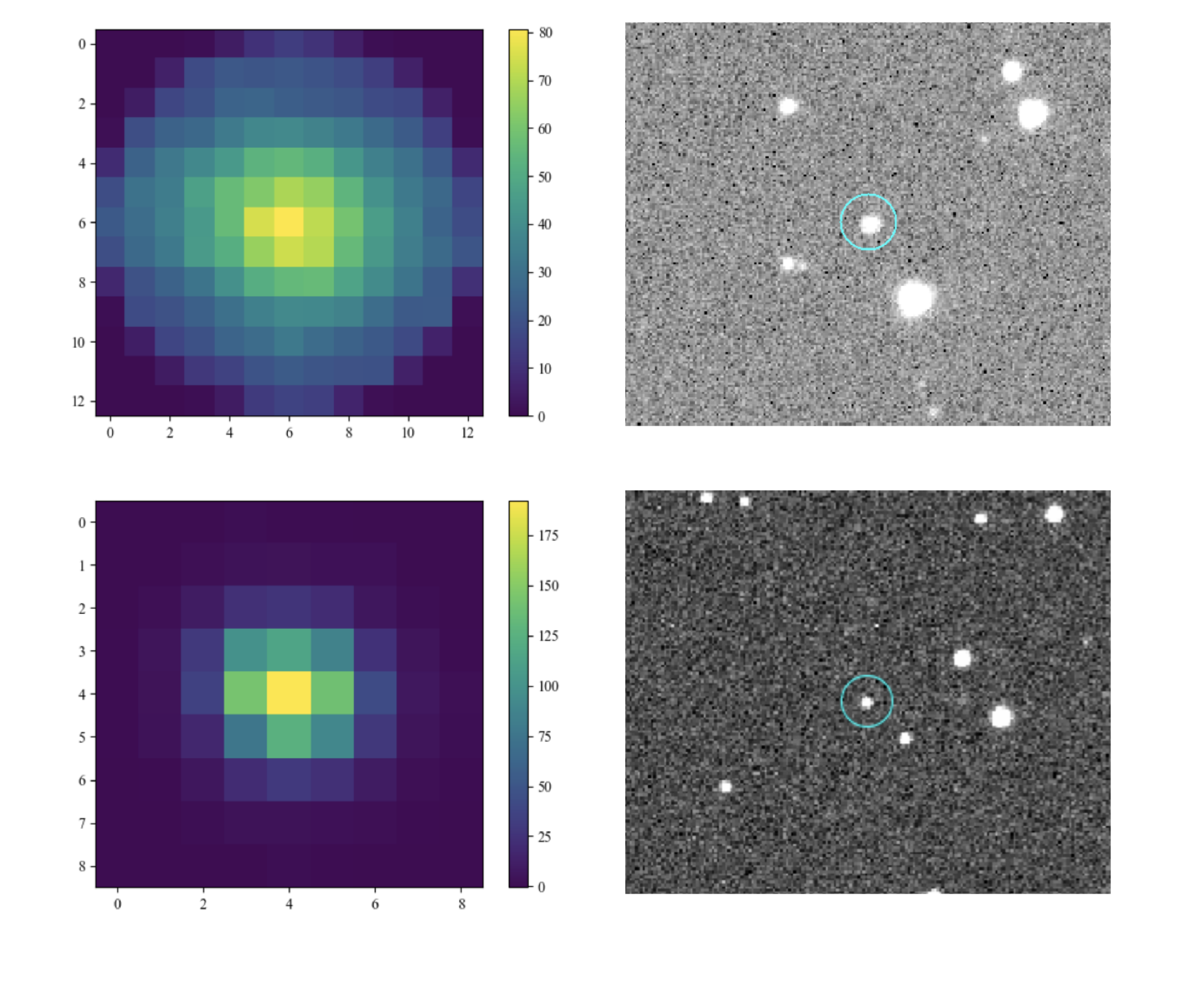}%
\caption{Examples of the PSF we inject into the image for DECam (top left) and iPTF (bottom left). The PSF is scaled to represent a source with $m_\mathrm{app}$ = 18 (left) and 17 (right), respectively. The median FWHM for this image is given as $\sim \ang{;;1.08}$, and $\ang{;;2.96}$, respectively. With a pixel scale of $\sim$0.26 arcsec\,pix$^{-1}$ for DECam and $\sim$1.01 arcsec\,pix$^{-1}$ for iPTF, we find a relative size of the injected PSFs to be $\sim$$\ang{;;3.38}$ and $\sim$$\ang{;;9.09}$, respectively. Representative samples of the injections in the DECam and iPTF data are also shown to the right of each respective model PSF.}
\label{fig:ppln_inj_sub_ex}
\end{figure*}

\subsection{Transient Injection Method}

The images we use are taken from public databases, previously processed and assumed to be free of unidentified transients. For iPTF and DECam, we have approximately 810 and 5650 individual CCD frames, respectively. We can then inject artificial transients in the images in order to test the efficiency and limitations of our recovery method. We limit the injection set to 50 per image per magnitude bin to avoid source confusion, with a total of approximately 40,500 and 282,500 injections per magnitude bin for iPTF and DECam images, respectively. We perform initial object detection, measure centroid positions, and determine aperture magnitudes on the template images using \texttt{SExtractor}.

Similar to ensemble building for Zernike statistics, we filter the observational image catalog for only isolated, high S/N and unsaturated sources with \texttt{CLASS\_STAR} close to 1. Identical to the Zernike method, we isolate the sources on a sufficiently large pixel grid, subpixel shift the centroid to the center of the grid, and approximate the PSF by averaging and triple sigma-filtering to remove outliers.

If subpixel shifting moves the centroid by more than 0.5\,pix, interpolation errors propagate and an inaccurate model will be produced. We therefore introduce a tiling procedure which discretizes the centroid pixel into a 3$\times$3 subgrid. Populations of sources are then grouped by the location of the centroid with respect to the minor grid coordinates. A separate PSF is then found by averaging over the ZP's for each section of the 3$\times$3 subgrid separately. For a randomized set of injection coordinates the appropriate PSF is chosen from the fractional pixel value of the centroid.

We coadd all sources together and average them, returning the error between each individual weighted source and the average PSF. To minimize the error, we apply triple sigma-clipping to throw out outliers and are finally left with a model transient to inject. 

We note that the same stars are selected to build the Zernike statistics as well as the model transients. With respect to the real sources in the image, it is possible in certain instances that the transients themselves may not be fully representative as a set. To ensure that our reported metrics are not biased, we perform visual checks of the median ensemble PSFs prior to injection as well as in each image after injection. Fig.\,~\ref{fig:ppln_inj_sub_ex} shows examples of what these injections look like in the sample DECam and iPTF images.

The field coordinates of the imaged region define a boundary for R.A. $\alpha$ and decl. $\delta$ and within the boundary we generate a random pair ($\alpha_i$, $\delta_i$) for each injection $i$. We randomize the flux values $f_i$ corresponding to a range of apparent magnitudes for sources which rise marginally above the background to sources that are short of reaching the CCD's nonlinear regime before becoming saturated. The normalized PSF is multiplied by $f_i$ and then subjected to a pixel-wise Poissonian distribution to simulate photonic shot noise.

To simulate the effects of variable observing conditions and optical quality degradation we generate additional background noise across the entire image and implement an optional blurring method, by which the entire image is convolved with a Gaussian kernel of variable width, see \S\ref{falsepos}. Fig.\,\ref{fig:ppln_inj_sub_ex} shows injection examples obtained for a set of both iPTF and DECam images.

As we have full control over the list of objects injected, we check for efficiency and FAR by cross-matching the coordinates of the transient candidate list and the observed image catalog, as well as the transient candidate list and the placed injections. We perform the cross-matching between catalogs using a descending KDTree algorithm. 

\subsection{Detection Efficiency}

\begin{figure}[t]
\includegraphics[width=\columnwidth]{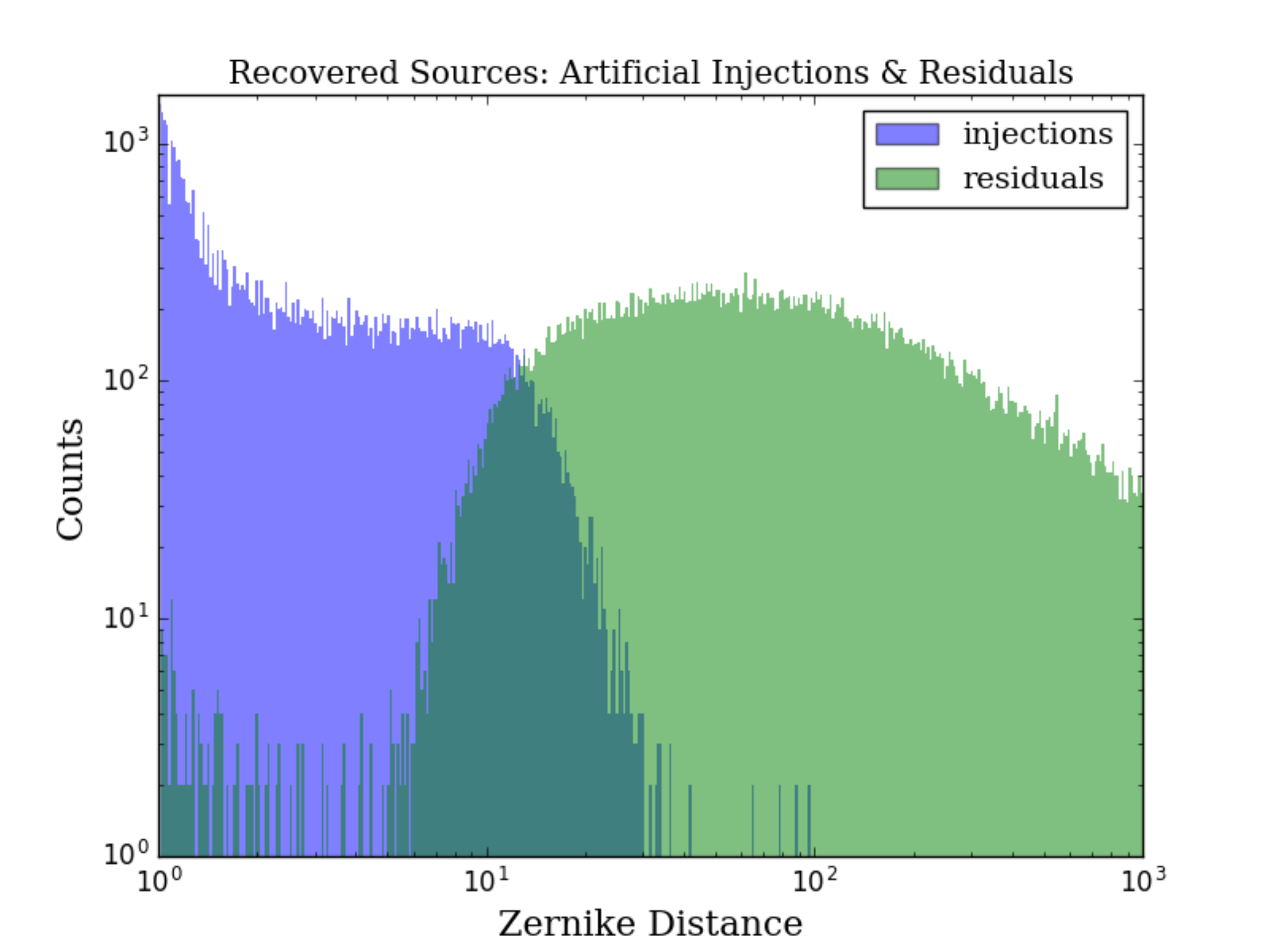}
\caption{Histogram of log-scaled recovered source counts versus log-scaled Zernike distance, with residuals coded in green and injections which span magnitude range of $16 \leq m_\mathrm{app}\leq 23$ coded blue for DECam. Note that the injected transients, which represent actual transients occupy a distinct Zernike space distinguishable from subtraction residuals.}
\label{fig:inj_art_hist_decam_bkgsn}
\end{figure}

\begin{figure}[t]
\includegraphics[width=\columnwidth]{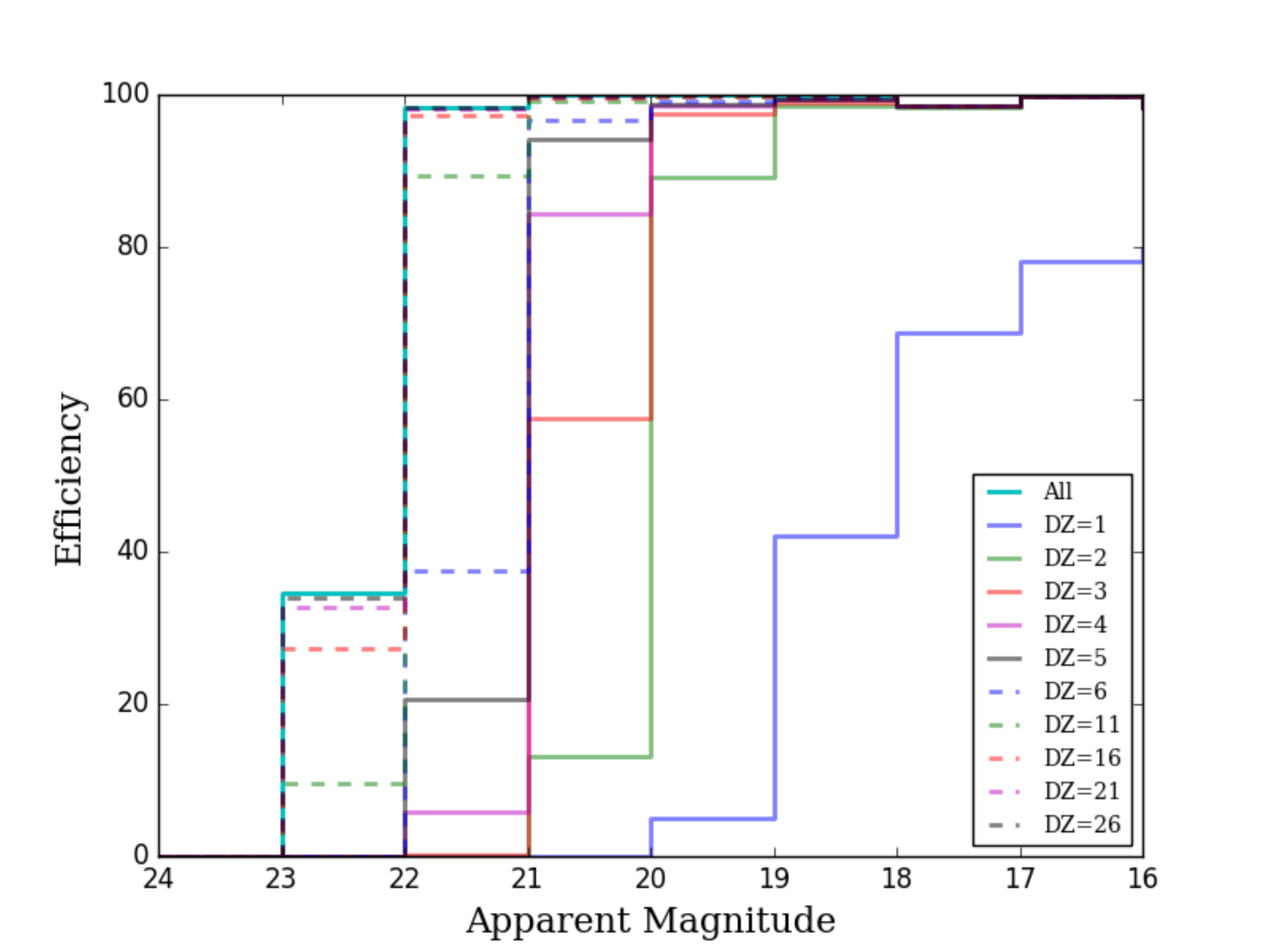}\\ \\
\includegraphics[width=\columnwidth]{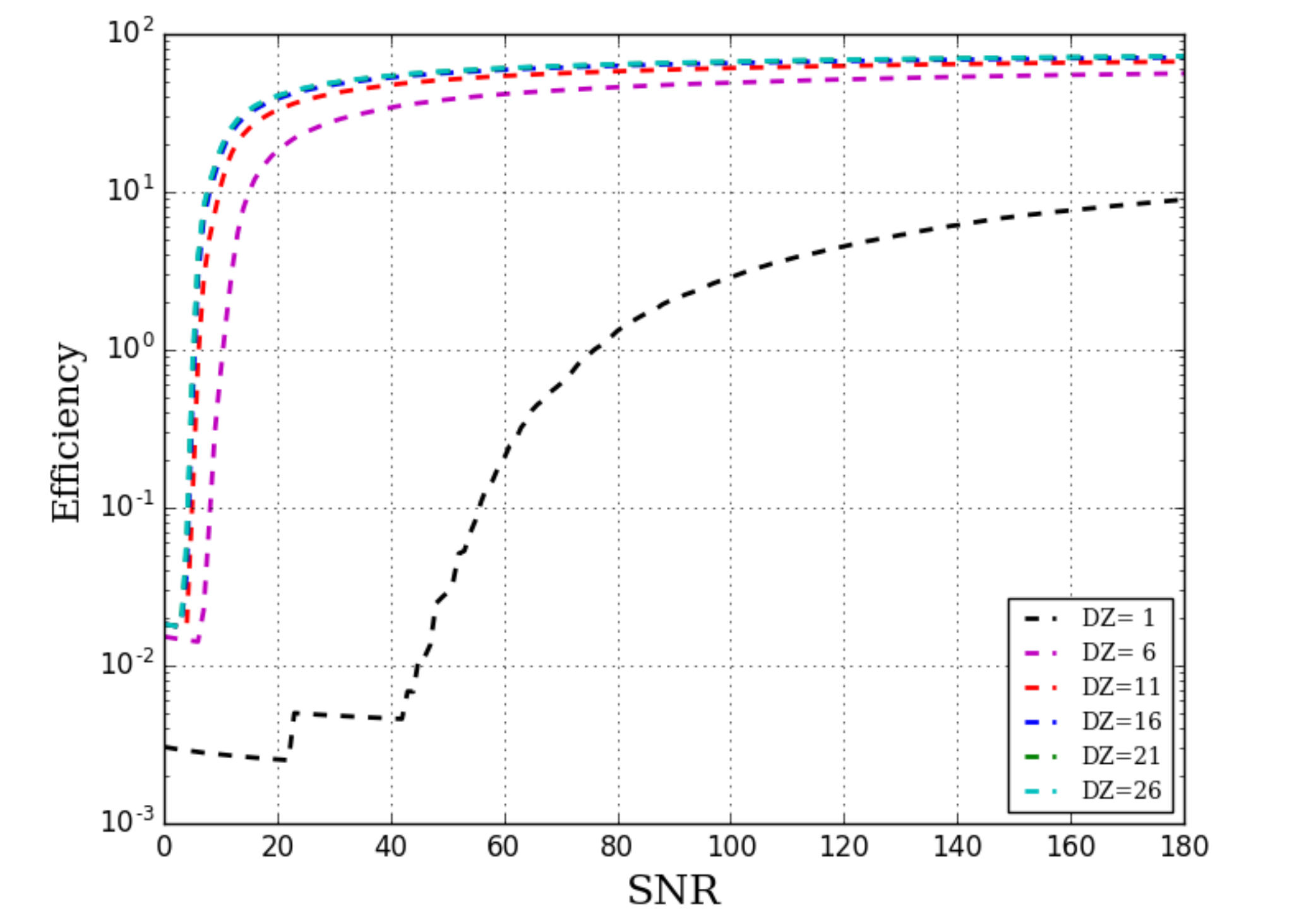}%
\caption{(a) Simulated detection efficiency versus apparent magnitude for DECam. By requiring a minimum detection efficiency of 90\%, we find an initial threshold of $D_Z = 11$ for sources with $m_\mathrm{app} \geq$ 22 for DECam and $D_Z = 11$ for sources with $m_\mathrm{app}\geq$ 18 for iPTF. The final threshold is tuned against both the efficiency and rate of false positives, however, the initial threshold serves as a first-pass estimate. (b) Efficiency versus SNR for DECam images for sources with apparent magnitudes brighter than $m=22$.}
\label{fig:eff_vs_mag}
\end{figure}

\begin{figure*}[t]
	\includegraphics[width=\textwidth]{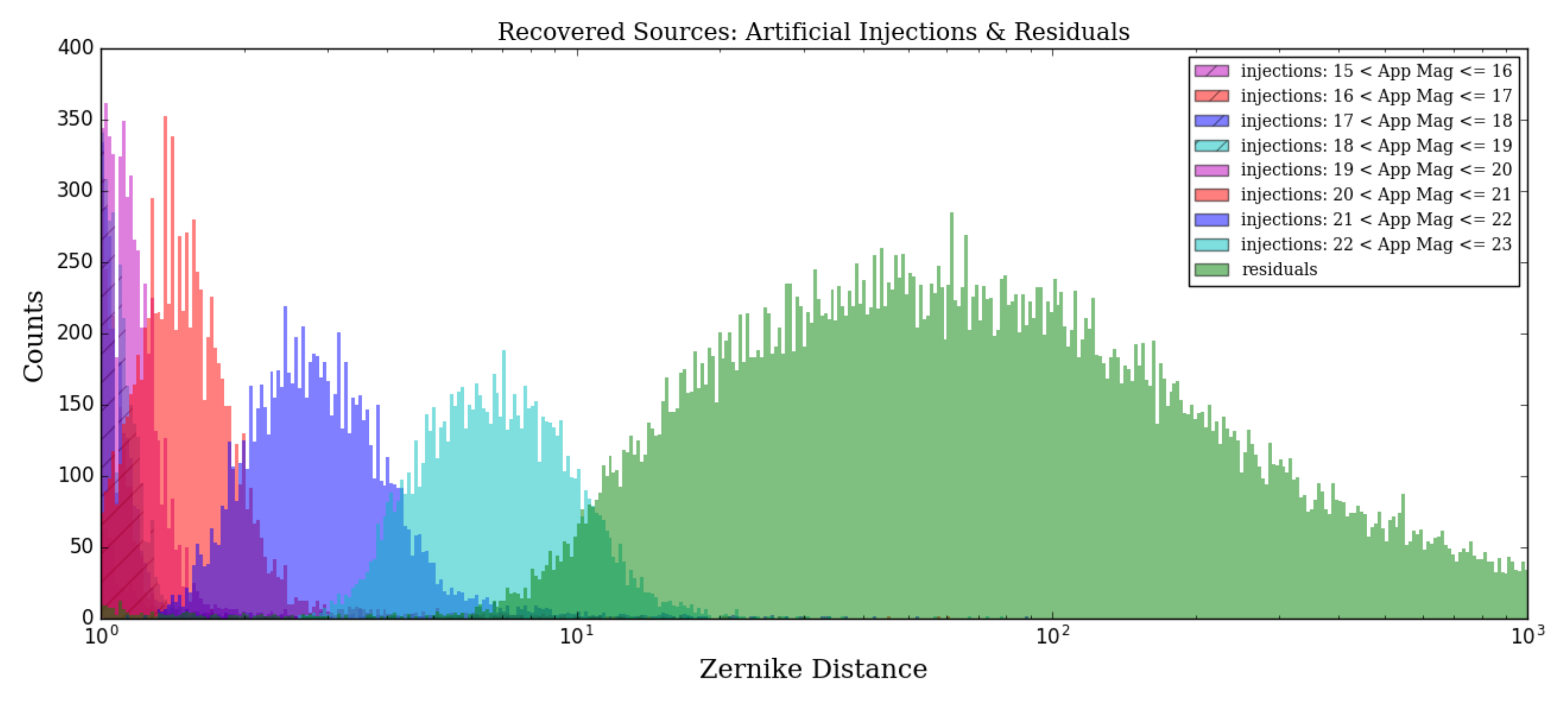}%
	\caption{Histogram of recovered sources both injected and residual objects in terms of their $D_Z$. We show the coverage of apparent magnitudes in Zernike space, where brighter sources will typically occupy lower regimes. As sources decrease in magnitude, and thus decrease in SNR, the separable nature of astrophysical transients and image artifacts becomes less concrete. This figure shows the potential of further tuning the threshold parameter once the apparent magnitude of a particular source is known.}
	\label{fig:hist_app_mag_sep}
\end{figure*}

For this paper, we disregard specific astrophysical model properties (brightness, light curve, etc.) and inject sources with random magnitudes within and just beyond the sensitivity of the detector. We apply the KDTree algorithm to recover our injections from the subtracted image transient list and record the candidate's $D_Z$ value. We can see the overall distributions of injections and artifacts/residuals in Fig.\,\ref{fig:inj_art_hist_decam_bkgsn}. We can distinguish two classes of image-subtracted objects represented as a bimodal distribution in Zernike space. We find an overall recovery efficiency of 98.64\% for DECam and 97.36\% for iPTF. In Fig.\,\ref{fig:eff_vs_mag} we show that we recover $>$99\% of our injections for sources with an apparent magnitude $m_\mathrm{app} <$ 22\,mag for DECam. Identification of faint sources in the subtracted image, i.e. $m_\mathrm{app} >$ 23 mag, proves more difficult. The SNR of the the faint sources is too low and as the relative impact of background noise increases, their spread in Zernike distance space increases. Fig.\,\ref{fig:eff_vs_mag} shows the recovery efficiency for injected objects with respect to (a) apparent magnitude and (b) SNR. By parameterizing over apparent magnitudes as a function of $D_Z$ and by requiring $>90$\% efficiency over a range of apparent magnitudes, we can estimate an initial $D_Z$ threshold. 
Further, Figs.\,\ref{fig:inj_art_hist_decam_bkgsn} and \ref{fig:hist_app_mag_sep} shows the separable nature of sources in Zernike distance space by their apparent magnitudes. A threshold $D_Z$ can then be further tuned based on knowledge of the signal strength and apparent magnitude of the transient for follow-up observations. While it is possible to subdivide the Zernike distance space in order to increase the sensitivity of a search for particular sources based on SNR and magnitudes, such considerations cannot be made without quantifying background rate of false positives (FPs) of our method, both under ideal and variable observing conditions.


\subsection{False Positives}
\label{falsepos}

\begin{figure}[t]
\includegraphics[width=\columnwidth]{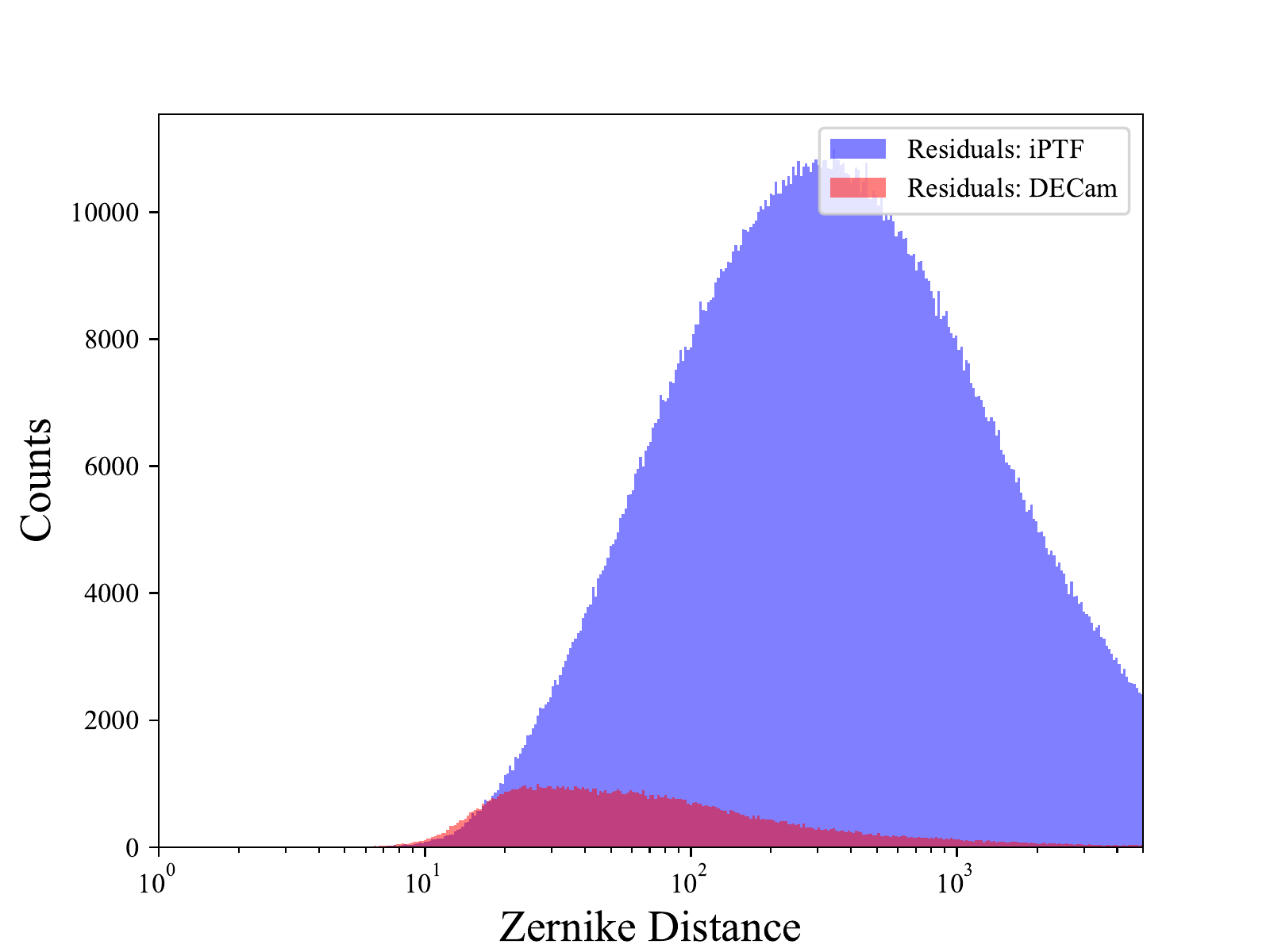}
	\caption{False positive count of the pipeline versus the log-scaled Zernike distance for iPTF (blue) and DECam (red). No sources are placed in the images and the image is subtracted against itself with pixel-wise background variations randomized about 1$\sigma$ of the normal background distribution.}
	\label{fig:far_vs_D}
\end{figure}

In our study here, FPs are defined as the non-astrophysical residuals in the subtracted image that pass all checks designed to discard them. To prevent true astrophysical transients incidentally present in these images from polluting the FP rate study, images are subtracted against themselves. Subtracting an image from itself mimics a perfect image subtraction routine and any potential astrophysical emission is entirely removed. Image registration, i.e. correcting for linear offsets and nonlinear warping across the frame, typically dominates the number of image subtraction residuals. Thus, in effect, we also remove the residuals which may appear from the registration process as there is no rotation or physical offsets between exposures to correct.  Additionally, any sources which remain would be true FPs of the image subtraction process itself. As the inherent design of the subtraction process is not adept at handling single-image subtraction (tending to lead to segmentation faults during execution), we apply artificial background variations to the science image. Fig.\,\ref{fig:far_vs_D} shows the distribution of FPs for iPTF and DECam in Zernike space. We can see from both distributions that the residuals occupy typical ranges of $D_Z \gtrsim 10$ under ideal conditions. 

Over a typical observing run, stochastic atmospheric turbulence will also deform the shape of the PSF and as the seeing worsens, the PSF spreads. To test the seeing-limited capabilities of our pipeline, we subtract a single image against itself holding the reference image constant and blurring the secondary, or observational, image with a Gaussian kernel, in effect mimicking stochastic processes from Kolmogorov turbulence in the atmosphere. In the continuous limit, blurring is approximated by convolution integrals. Since we apply our method in the discrete regime and because the PSFs are spatially invariant, the convolution is a linear process. Therefore we replace the integral with a sum,
\begin{equation}
I_b(\alpha,\delta) = \sum_{a=-\infty}^{+\infty}\sum_{b=-\infty}^{+\infty}I_o(\alpha+a, \delta+b)K_X(-a,-b),
\end{equation} 
where $I_o(\alpha,\delta)$ is the observational image and $K_X(-a,-b)$ is the discrete convolution kernel. 

\begin{figure}[t]
	\includegraphics[width=\columnwidth]{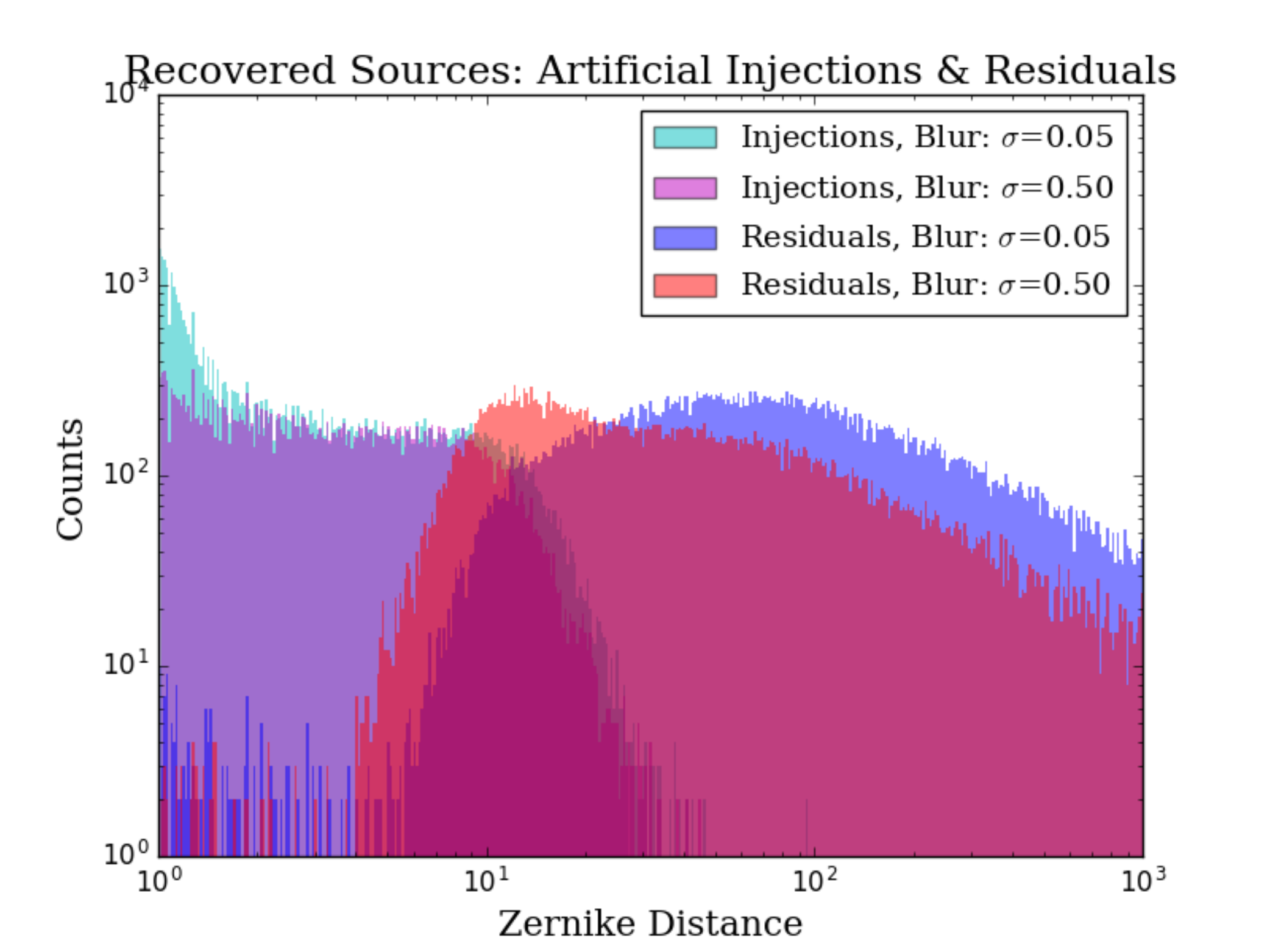}%
	\caption{Comparison of minimal to maximal blurring effects on the Zernike distance of recovered sources. The degree to which the image is blurred affects the rate of false positives.}
	\label{fig:blur}
\end{figure}

\begin{figure}[t]
	\includegraphics[width=\columnwidth]{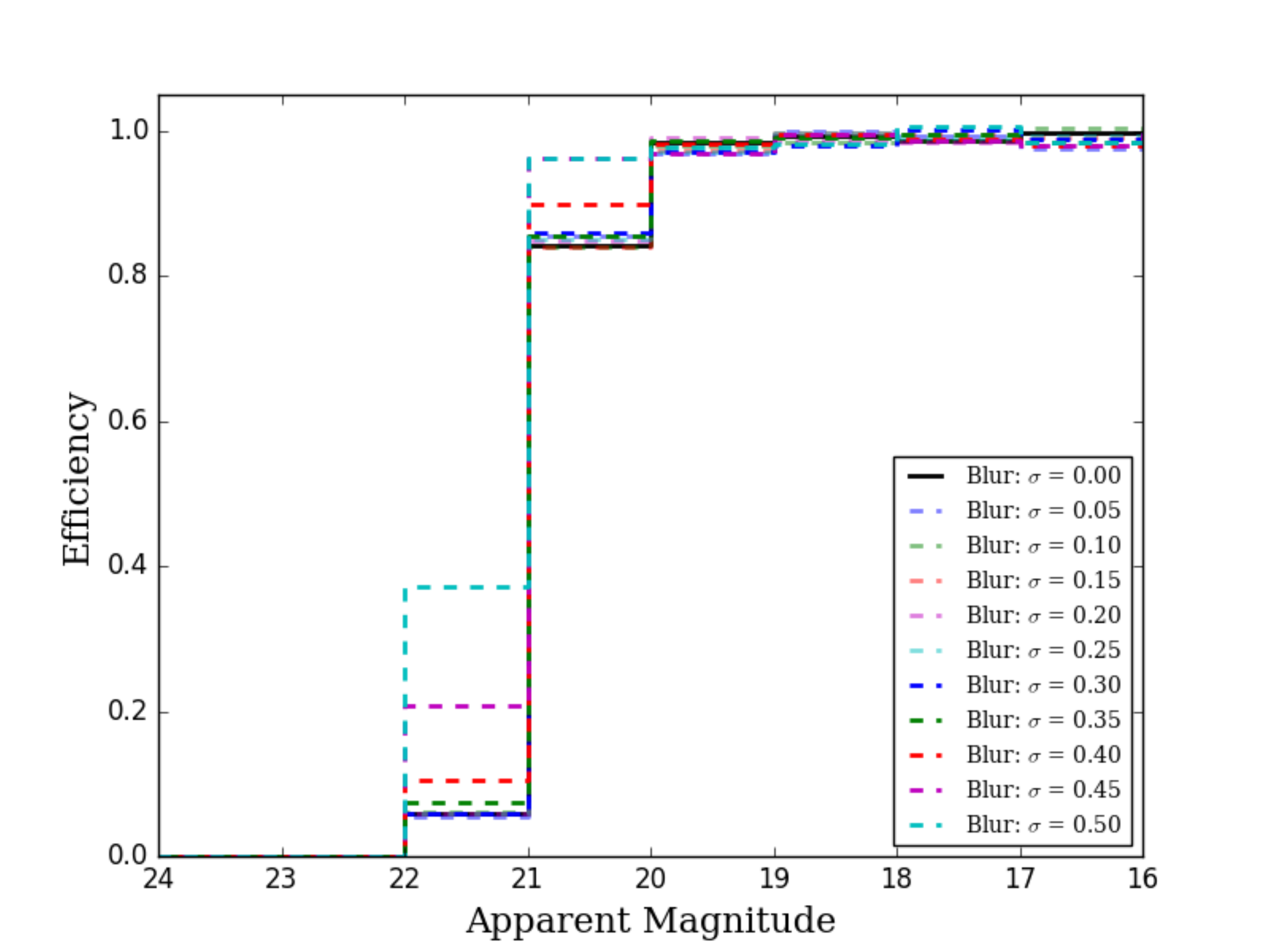}%
	\caption{Efficiency versus apparent magnitude under variable seeing conditions for DECam images with a $D_Z$ = 4 threshold and Gaussian kernels ranging from assumed idealized seeing to maximal blurring.}
	\label{fig:eff_vs_mag_blur}
\end{figure}

The Gaussian convolution kernel is defined in the standard way, where
\begin{equation}
K_{G}\left(x,y\right) = \frac{1}{\sqrt{\pi\sigma_G^{2}}}e^{-\frac{x^2+y^2}{\sigma_G^2}}.
\end{equation}
We vary $\sigma_G$ in step sizes of 0.05\,pix between 0.05 and 0.5\,pix, where values larger than 0.5\,pix express nonlinear representations of optical blurring. Fig.\,\ref{fig:blur} shows the effects of simulated suboptimal observing conditions on our ability to distinguish artifacts from true transients as a function of $D_Z$. We see in Fig.\,\ref{fig:blur} that we recover about an order of magnitude fewer sources at low Zernike distances, i.e. $D_Z \sim 1$, and that the distribution of residuals is pushed towards lower $D_Z$, becoming less distinguished from our set of injections. The PSF of an image may be approximated by a Gaussian. However, in general, it is not a true Gaussian. Thus, when the image is convolved with a Gaussian kernel the PSF takes on Gaussian properties and tends to smooth each object, shifting neighboring counts toward the centroid position of the PSF and increasing the SNR in the process. In effect, it provides better sampling under worse seeing with respect to matching the PSF to the set of ZPs. There is also the effect that the noise in the Zernike distance will be smoothed out and that this effect increases with decreasing brightness of sources. Additionally, Fig.\,\ref{fig:eff_vs_mag_blur} shows that as we approach the lower magnitudes our efficiency increases, as we effectively render faint sources more point-like. However, this comes at a cost as we simultaneously increase the total number of objects which pass through, particularly FPs, as image subtraction residuals are also rendered more point-like. For brighter magnitudes we do not see a significant response to our efficiency under the effects of blurring as already high SNR sources are made even more point-like.

\subsection{Receiver-operating Characteristic Curve}
\begin{figure}[t]
\includegraphics[width=\columnwidth]{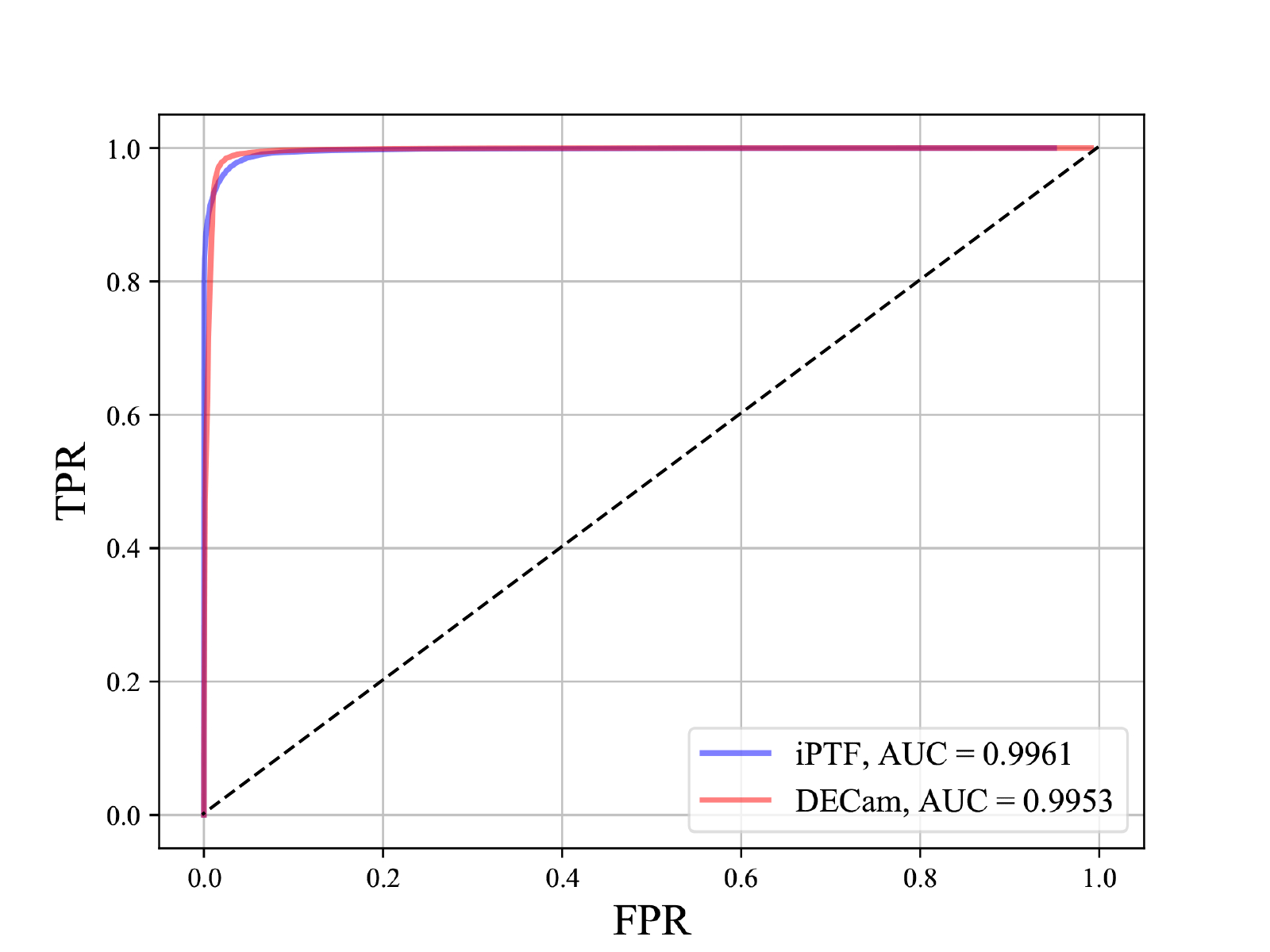}
\caption{ROC Curve for iPTF (blue) and DECam (red). With a maximum false positive rate of 1\%, we find an true positive rate of 92.8\% and 91.5\% respectively. For a random classifier AUC=0.5 and is represented as the black dashed diagonal line.}
\label{fig:roc_decam_bkgsn}
\end{figure}

A truly automated system has minimal human intervention, thus we must leverage a threshold $D_Z$ against both the number of FPs which pass our checks and the efficiency of recovery. ROC curves\,\citep{Fawcett2006} are performance predictors for binary classification where the boundaries between classes are determined by a typical threshold value, e.g. $D_Z$. For a threshold value of $D_Z=4$ (6) for DECam (iPTF) and a maximum FPR=1\% in Fig.\,\ref{fig:roc_decam_bkgsn} we find a true positive rate (TPR), or efficiency, of 91.5\% (92.8\%). TPR and FPR are defined as ratios of the numbers of true positives (TP), FPs, true negatives (TN) and false negatives (FN), or 
\begin{eqnarray}
\mathrm{TPR = \frac{TP}{TP+FN}} \\
\mathrm{FPR = \frac{FP}{FP+TN}},
\end{eqnarray}
respectively.

The ROC curve presents a way to analyze our choice for an appropriate $D_Z$ threshold. In terms of binary classification, the area under the ROC curve (AUC) defines the performance of our model with respect to our discriminator. For a suboptimal model, i.e. the $D_Z$ threshold set too low or too high, we might find our distributions indistinguishable and AUC=0.5. The higher the AUC value, the better performance our model has at distinguishing between true and false positives.

\section{Discussion}
\label{discussion}

We compare the relative performance of our automated method to the supervised methods used by the iPTF and DES-SN surveys. Both surveys use an RF algorithm to discriminate between real and bogus sources and recover 92.3\% and 88\% of TPs, respectively. We find relatively comparable results using our method, or a recovery of 92.8\% and 91.5\% of TPs, respectively. We compare our results to additional alternative machine-learning algorithms in Table~\ref{tab:ml_comp}. We note that we cannot make direct comparisons between our work and those of other data sets and algorithms. Any comparisons we report are purely relative as we have not attempted to standardize the data sets in any way.

The simulations we perform take into account a variety of observing scenarios over a realistic sample of full-sky measurements. We account for the differences in inherent image noise and changes in observing conditions, primarily seeing, by adding random pixel-wise shot noise and blurring the image with a Gaussian kernel, respectively. The addition of shot noise is not a significant contribution to the overall results, and thus is not expanded upon in great detail. Under full-image blurring, we see that the fraction of recovery increases for faint sources, i.e. $m_\mathrm{app}\sim 22$, by a factor of 8, while the efficiency of recovery for brighter sources remains consistent. However, the efficiency increases for faint sources simultaneously increases the rate of FPs. As we Gaussianize the injections artificially via blurring the entire image, we also do so to the other sources we wish to remove, i.e. leaving behind more characteristically point-like objects. Our method is able to recover artificially injected transients under extreme conditions, although the rate of FPs will correspondingly increase.

We introduce the AUC metric to quantify the discriminatory power of our model for binary classification. We vary our discriminator $D_Z$ and calculate the AUC for DECam and iPTF, given in Table~\ref{tab:effauc}, and show that under all simulated conditions AUC $>0.99$.

\subsection{Run Time}
\label{runtime}
The software is built in Python and will be released as an open-source product. Run times were computed using a Intel i7-3770k 8-core chip clocking at 3.50GHz. Parallelizations have been implemented using the standard Python \texttt{multiprocessing} module. The following run times are given for single CCD images and are calculated for the image subtraction stage, final \texttt{SExtractor} task, Zernike distance calculation, and I/O operations. The average (median) user run times for iPTF and DECam are 394 and 80\,s (186 and 68\,s), respectively. Processor time on the other hand was more efficient with an average (median) of 32 and 38\,s (29 and 35\,s) for iPTF and DECam, respectively. The user and CPU run times are processor-dependent and the major contribution to timing differences are related to I/O operations. However, this provides a useful benchmark for the run times given the relative size of the CCD. Multiple cores and parallelization are necessary for the latency requirements of most fast-transient search cadences.

\section{Conclusions}
\label{conclusions}

We have presented a method of image analysis using a set of ZPs to identify and classify sources which are point-like and characteristically similar to expected astrophysical transients. We show that the Zernike distance is a tunable metric for each telescope and can be optimized to maximize the detection probability and minimize the FAR.

We reduce the overall number of residuals that must be manually vetted, which may appear as a direct consequence to image subtraction, based on shapelet analysis. We show that our method recovers $>99\%$ of real transients on archival single-pass observations from iPTF and DECam surveys and that our results are comparable to transient classification techniques that use supervised and unsupervised machine learning. Thus, we show that it is possible to identify potentially interesting transient objects while significantly reducing or entirely eliminating the total number of human hours needed to train the data. We must note that while SVD can be used as a method for dimensionality reduction in the context of unsupervised machine learning, we make use of SVD for polynomial decomposition.

We show that our method is robust under a variety of variable observing conditions, however, the increase in the number of FPs must be leveraged against nightly seeing.

Our results are also presented using only shape cuts in Zernike space and do not apply additional vetting. Sources which pass our criteria are sent on for human vetting, where we have bundled an automated interactive and customizable webpage for the user to analyze the source further and/or optionally send off automated alerts.

Future work may employ a univariate Gaussian mixture model to determine the probability of a particular source belonging to one of the classes. Additionally, future work includes increasing the dimensionality of our space to incorporate the Zernike distance as a function of secondary or tertiary observables, such as SNR, and using a multivariate Gaussian mixture model to cluster the parameter space.

Any transients found in an image, followed-up, and deemed interesting may be used for targeted GW searches. Under certain observing scenarios we can constrain and approximate the time of the transient EM event to search back into the GW data stream. As the distance to which a telescope can observe a signal is larger than the nominal GW detector distance horizon, looking for GW signals triggered by EM events may also lead to promoting subthreshold GW events; in effect increasing GW detectors' distance horizon.

\section{Acknowledgements}
We thank the anonymous referee for the extremely useful comments. We thank Jade Powell for comments on the manuscript. K.A. is supported by the Australian Research Council (ARC) Centre of Excellence for Gravitational Wave Discovery (OzGrav), through project number CE170100004. S.S.E. was supported in part by the National Science Foundation under award PHY-1806651. S.K. is supported by PHY1505308. A.G. funding is provided though NASA grants NNX15AL14G and NNX17AE11G LIGO was constructed by the California Institute of Technology and Massachusetts Institute of Technology with funding from the National Science Foundation and operates under cooperative agreement PHY-0757058. This manuscript has LIGO Document ID LIGO-P1800327. This work makes use of the open data of intermediate Palomar Transient Factory and Dark Energy Camera -- Dark Energy Survey Year 1.

% \software{Astropy \citep{2013A&A...558A..33A}, SciPy \citep{jones_scipy_2001}, NumPy \citep{van2011numpy}, matplotlib \citep{Hunter:2007}, ds9 \url{http://ds9.si.edu/site/Home.html}}

\bibliographystyle{apj}
\bibliography{zernike_pipeline}

\appendix

\begin{deluxetable*}{cccccc}[h]
\tablenum{1}
\tablecaption{Relative performance comparison of this work to typical machine-learning algorithms\label{tab:ml_comp}}
\tablewidth{0pt}
\tablehead{
\colhead{Survey} & \colhead{Author} & \colhead{ML Model} & \colhead{TPR (\%)} & \colhead{FPR (\%)} & \colhead{MDR (\%)}}

\startdata
Supervised Algorithms \\
\hline
PTF & \cite{Bloom2012,Brink2013} & RF & 92.3 & 1 & 7.7\\
Pan-STARRS1 & \cite{Wright2015} & RF & 90 & 1 & 10\\
DES-SN 	& \cite{Goldstein2015} & RF & 88 (96) & 1 (2.5) & 12 (4) \\
SDSS 	& \cite{duBuisson2015} & \textbf{RF}, kNN, SkyNet, nB, kSVM 	& 91 & 8.65 & 8.81 \\
Nearby SNFactory & \cite{Bailey2007} & \textbf{BDT}, RF, SVM & 95 & 1 & 5\\
HSC & \cite{Morii2016} & \textbf{AUC Boost, RF, DNN} & 90 & 1 & 10\\
\hline
Unsupervised Algorithms \\
\hline
OGLE-IV & \cite{Klencki2016} & SOM & 86 & 1 & 14 \\
\hline
\textbf{Shapelet Analysis}\\
\hline
DECam & This work & SVD & 91.5 & 1 & 8.5 \\
iPTF & This work & SVD & 92.8 & 1 & 7.2 \\
\hline \\
\enddata
\tablecomments{Relative comparison to typical machine-learning algorithms in terms of the true positive rate (TPR), the false positive rate (FPR), and the missed detection rate (MDR), or false negative rate. In the instances where there are multiple machine-learning models given, the models in bold are found to be the best performing. In the case for the HSC survey, a combination of all three models shows the best overall performance for their data set.}
\end{deluxetable*}

\begin{deluxetable*}{ccccccccccccccc}
\tablenum{2}
\tablecaption{Detection Efficiency\label{tab:effauc}}
\tablewidth{0pt}
\tablehead{
\colhead{Instrument} & \colhead{$\sigma_\mathrm{blur}$} & \colhead{Overall Efficiency (\%)} & \multicolumn{7}{c}{Efficiency (\%) for Apparent Magnitude} & \colhead{AUC} \\
\cline{4-9}
			& & & 16&17&18&19&20&21 & }

\startdata
DECam 	&0.00&97.36&98.92&98.81&99.01&98.67&97.31&91.42&&0.9953 \\ 
		&0.05&98.31&99.12&99.33&99.18&99.05&98.28&94.93&&0.9977 \\ 
		&0.10&98.35&99.19&99.41&99.14&99.13&98.29&94.91&&0.9974 \\ 
		&0.15&98.37&99.28&99.20&99.38&98.94&98.40&95.02&&0.9981 \\ 
		&0.20&98.36&99.25&99.36&99.17&99.04&98.50&94.85&&0.9977 \\ 
		&0.25&98.32&99.06&99.10&99.12&99.17&98.41&95.05&&0.9985 \\ 
		&0.30&98.39&99.39&99.28&99.34&99.04&98.31&95.00&&0.9976 \\ 
		&0.35&98.40&99.20&99.36&99.28&99.20&99.40&94.99&&0.9973 \\ 
		&0.40&98.43&99.20&99.48&99.24&99.19&98.58&94.89&&0.9979 \\ 
		&0.45&98.52&99.39&99.43&99.44&99.10&98.69&95.11&&0.9979 \\ 
		&0.50&98.73&99.68&99.50&99.54&99.35&98.71&95.62&&0.9977 \\
\hline
		& & & 14&15&16&17&18&&&& \\
\hline
iPTF 	& 0.00&98.64&99.08&99.03&98.99&97.47&73.08&&& 0.9938 \\ 
		& 0.05&98.42&98.79&98.80&99.15&96.96&69.63&&& 0.9957 \\ 
		& 0.10&98.77&99.00&99.67&98.94&97.48&71.88&&& 0.9956 \\ 
		& 0.15&98.80&98.88&99.25&99.06&98.04&72.03&&& 0.9954 \\ 
		& 0.20&98.49&98.93&98.99&99.21&96.82&73.88&&& 0.9957 \\ 
		& 0.25&98.94&99.22&99.43&99.84&99.27&73.19&&& 0.9953 \\ 
		& 0.30&98.79&99.36&99.64&99.07&97.09&71.46&&& 0.9955 \\ 
		& 0.35&98.38&99.15&99.03&98.57&96.77&69.69&&& 0.9950 \\ 
		& 0.40&98.09&98.59&98.87&99.07&95.84&68.52&&& 0.9932 \\ 
		& 0.45&98.01&98.93&99.25&98.38&95.50&67.84&&& 0.9916 \\ 
		& 0.50&98.48&99.57&99.40&99.05&95.89&69.35&&& 0.9912 \\ 
\enddata
\tablecomments{Detection efficiency for sources parameterized by apparent magnitudes and by the amount of generated blurring. We blur the images to simulate variable observing conditions, where the larger the $\sigma_{\mathrm{blur}}$, the worse the estimated seeing. We recover $\gtrsim 99$\% of sources brighter than $m>21$ for DECam and $\gtrsim 96$\% down to $m>18$ for iPTF with an overall efficiency of $\gtrsim 98$\% for both instruments. We report the AUC for each model which indicates the appropriateness of our model given the data.}
\end{deluxetable*}

\end{document}